# On the synthesis methodologies to prepare 'Pb$_9$Cu(PO$_4$)$_6$O': phase, composition, magnetic analysis and absence of superconductivity

Gohil S. Thakur,[a,b*] Manuel Schulze[a] and Michael Ruck.[a,c*]

[a]Technische Universität Dresden 01062 Dresden, Germany

[b]Würzburg-Dresden Cluster of Excellence, *ct.qmat*, Technische Universität 01062 Dresden Dresden, Germany

[c]Max Planck Institute for Chemical Physics of Solids Nöthnitzer Str. 40, 01187 Dresden, Germany

**Abstract**

We present the results of our various attempts to prepare the presumably room-temperature and ambient-pressure superconducting compound 'Pb$_9$Cu(PO$_4$)$_6$O' (LK-99). We experimented with various starting materials, and used several synthesis techniques like, sealed quartz tubes and air-sintering pathways to prepare the reported phase. Repetition of the exact synthesis process followed by Lee *et al* failed to reproduce the superconducting phase yielding only a multiphase sample. None of our prepared samples exhibits Meissner effect or Levitation. Very importantly, no copper (or only a trace amount) could be detected in any of the designated 'LK-99 phase' with PXRD pattern identical to the reported one! Additionally, some dark greyish flakes in some samples responded towards the magnet in a way imitating 'half-levitation'. The magnetic measurement on those sample suggest it is diamagnetic in the measured temperature range of 2-325 K but show a weak soft ferromagnetic behavior at 2 K, the origin of which is still unknown to us.

*Corresponding Author(s) gohil_singh.thakur@tu-dresden.de, Michael.ruck@tu-dresden.de

**Introduction**

A very recent claim of room-temperature superconductivity (RTSC) at ambient pressure by a group of Korean researchers on 22$^{nd}$ July 2023, sparked a new wave of interest in investigating ambient pressure RTSC above 300 K.[1,2] The group claimed to have achieved this in a Pb-oxyapatite structure $Pb_{10-x}Cu_x(PO_4)_6O$ (0.9 < $x$ < 1.1). Superconductivity (SC) at an astonishingly high critical temperature ($T_c$) of < 400 K is observed after substituting Cu in the known parent insulating compound $Pb_{10}Cu(PO_4)_6O$[3], where Cu supposedly occupies some of the Pb2 sites. The scientific community immediately took upon the task of verifying this claim and made several attempts to replicate the observation of RTSC.[4-9] However, results have not been convincingly reproduced and verified yet. In this article, we present the results of our elaborate efforts to reproduce the sample prepared by Lee *et al*[1,2] and verify its superconducting properties. The sample by Lee *et al* [1,2] was prepared in a unique way by a metathesis reaction of equimolar mixture of $Pb_2OSO_4$ (mineral Lanarkite) with $Cu_3P$ in a sealed quartz tube at high temperature to yield $Pb_9Cu(PO_4)_6$. However, since non-stoichiometric amounts of reactants were used a lot of side phases also resulted along with the desired phase. Until now, there is no experimental procedure that would yield a single-phase material. Hence, it is very important to find a method to obtain this LK-99 phase as a phase pure material and perform detailed characterization on it in order to understand the reason for such a high $T_c$. A phase pure materials would also help is establishing the correct crystal structure of the material which is essentially required for performing quantum chemical calculation to understand the superconducting properties of this material. Additionally, Wu *et al*[10] observed very interesting transitions at ~380 K in both resistivity as well as in magnetic susceptibility in pure $Cu_2S$ and LK-99 containing significant $Cu_2S$ phase, which exactly mimics the (superconducting) transition observed by Lee *et al* in LK-

99. Thus, it is of utmost priority to produce a very clean sample, especially free form $Cu_2S$ impurity, to determine the intrinsic properties of LK-99 and to conclude whether the superconducting transition actually originates from the $Pb_{10-x}Cu_x(PO_4)_6O$ phase. We conducted detailed experiments to reproduce the compound by the original reported route and in addition to several new alternative routes, which should theoretically yield no side products. Most of our modified synthetic routes however, did not yield the LK-99 phase. In few experiments, we achieved the seemingly same material as reported, but no SC was not observed in them. With the help of detailed compositional analysis, we highlight the reluctance of copper to be incorporated at the Pb-sites, which supposedly results in the absence of superconductivity. This work also outlines the difficulties in preparing a homogeneous phase-pure material, which otherwise appear simple in theory.

**Experimental:**

**Synthesis.** Synthesis of the $Pb_{10-x}Cu_x(PO_4)_6O$ was carried out using several routes listed below. We used evacuated sealed quartz tube as well as air-sintering to prepare the samples. Experiments I – V were performed in evacuated quartz tubes at 925 °C with dwell time of 10-20 h and the rest (VI-VII) were performed in an air atmosphere in platinum crucibles under the same heating conditions. The precursors $Pb_2OSO_4$, $Cu_3P$ were prepared exactly as reported by Lee *et al*[1]. CuO (abcr, 99.99 %), Cu powder (Sigma Aldrich, 99.9, reduced in $H_2$ stream at 400°C before use) $P_2O_5$ (Riedel-de Haën, 98 %), PbO (abcr, 99.999%) and $(NH_4)_2HPO_4$ (Thermoscientific, 98%) were used as purchased after an initial check for phase purity by PXRD. Monoclinic $Pb_3(PO_4)_2$ and triclinic $Cu_3(PO_4)_2$ were prepared by direct reaction of stoichiometric $(NH_4)_2HPO_4$ with PbO and CuO, respectively. The starting mixtures were kept in Pt crucibles and heated in air at 900°C for 20 h. All the prepared starting materials were checked for their phase purity before use. Except for

Cu$_3$P (which had an evident 10 % unreacted Cu metal impurity) all the raw materials used were X-ray phase pure.

I. $2\ Pb_2OSO_4 + 2\ Cu_3P \rightarrow 1/3\ Pb_9Cu(PO_4)_6O + 2\ Cu_2S + 5/3\ CuO + Pb$

II. $8\ PbSO_4 + PbO + 3Cu + 6\ P_{red} \rightarrow Pb_9Cu(PO_4)_6O + 1\ Cu_2S + SO_2\uparrow + 3\ O_2\uparrow$

III. $6Pb_2OSO_4 + CuO + 6\ P_{red} \rightarrow Pb_9Cu(PO_4)_6O + 3\ PbS + 3\ O_2\uparrow$

IV. $9\ PbO + CuO + 3\ P_2O_5 \rightarrow Pb_9Cu(PO_4)_6O$

V. $3\ Pb_3(PO_4)_2 + CuO \rightarrow Pb_9Cu(PO_4)_6O$

VI. $8\ Pb_3(PO_4)_2 + Cu_3(PO_4)_2 + 3\ PbO \rightarrow 3\ Pb_9Cu(PO_4)_6O$

VII. $9\ PbO + CuO + 6\ NH_4H_2PO_4 \rightarrow Pb_9Cu(PO_4)_6O + 6\ NH_3\uparrow + 9\ H_2\uparrow$

**Diffraction studies.** Room temperature powder X-ray diffraction (PXRD) data (CuK$\alpha_1$, $\lambda$ = 154.059 pm, $T$ = 296(1) K) were collected using an X'Pert Pro diffractometer (PANalytical, Bragg-Brentano geometry, curved Ge-(111) monochromator, fixed divergence slits, PIXcel detector). The samples for the PXRD measurements were ground and fixed on a single-crystal silicon sample holder.

Semi-quantitative elemental analysis of the samples was performed using Scanning electron microscopy equipped with Energy dispersive Analysis by X-rays (SEM-EDAX, SU8020 (Hitachi), equipped with a Silicon Drift Detector (SDD) X-MaxN (Oxford)).

**Magnetic Measurements.** Magnetization data was collected using a vibrating sample magnetometer (VSM) option on a 9T cryogenic free measurement system (CFMS, Cryogenics Ltd.). Variable temperature magnetization was collected in the temperature range 2-325 K in an applied magnetic field of 50 Oe in zero field cooled (ZFC) and field cooled conditions (FC). Data was collected in warming runs. Isothermal magnetization at $T$ = 2 and 325 K was recorded in an applied field of ±1 T.

**Result and discussion**

'LK-99' or $Pb_9Cu(PO_4)_6O$ crystallizes in the Pb-apatite family of structures with hexagonal crystal system. It is a doped derivative of the $Pb_{10}(PO_4)_6O$, also known as oxypyromorphite, (Figure 1) where Pb1 and Pb2 occupies a nine coordinated sites each and O atoms form a near regular tetrahedral around P atoms. Only one additional oxygen is not the part of $PO_4$ tetrahedra and is only 25 % occupied. The substituted copper in $Pb_9Cu(PO_4)_6O$ supposedly sits at the Pb2 sites.

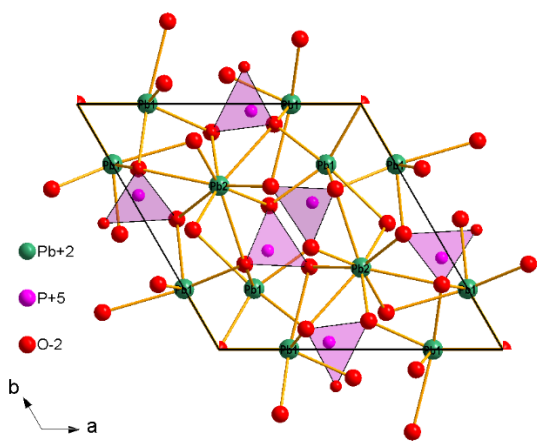

Figure 1. Crystal structure of $Pb_{10}(PO_4)_6O$ viewed along the crystallographic *c*-axis.

Table 1. Details of the synthesis routes adapted for preparation of LK-99 phase

| Expt. | Starting materials | PXRD | SEM-EDX | Remarks |
|---|---|---|---|---|
| I | $Pb_2(SO_4)O + Cu_3P$ | $Pb_{10}(PO_4)_6O$ type phase + $Cu_2S$ impurity | No or only trace copper detected in the sample | Diamagnetic but no levitation |
| II | $PbSO_4 + PbO + Cu + P$ | $Pb_{10}(PO_4)_6O$ type phase + small impurity | No copper detected in the sample | No levitation |
| III | $Pb_2SO_5 + CuO + P$ | $Pb_{10}(PO_4)_6O$ type phase (nearly pure), trace Cu impurity. PbS crystals on cold end of the tube. | No or only trace copper detected in the sample, regions of PbS and $Cu_2S$ also present | Diamagnetic but no levitation |
| IV | $PbO + CuO + P_2O_5$ | monoclinic $Pb_3(PO_4)_2$ phase | Pb:P = 3:2, No Copper detected in crystals | No levitation |

| V | Pb$_3$(PO$_4$)$_2$ + CuO | monoclinic Pb$_3$(PO$_4$)$_2$ phase | | Grayish powder |
|---|---|---|---|---|
| VI | Pb$_3$(PO$_4$)$_2$ + Cu$_3$(PO$_4$)$_2$ + PbO (Pt crucible) | Pb$_9$(PO$_4$)$_6$ and Pb$_3$Cu$_3$(PO$_4$)$_4$ phases | Pb:P = 1.6 and Cu rich phase (presumably Pb$_3$Cu$_3$(PO$_4$)$_4$ | Transparent greenish/yellow crystals |
| VII | PbO + CuO + (NH$_4$)$_2$HPO$_4$ (Pt crucible) | Mostly Pb$_8$O$_5$(PO$_4$)$_2$ phase | No Cu detected Pb:P ~ 4 | Transparent yellowish crystals |

*Phase analysis:*

Details of all the synthesis routes adapted for preparation of LK-99 phase and theier subsequent results are tabulated in Table 1. The PXRD patterns of all the Pb$_{10}$(PO$_4$)$_6$O type phase obtained are shown in Figure 2. These three experimental patterns match very well with the pattern of Pb-oxyapatite calculated from ref. [3] and with the LK-99 phase reported by Lee *et al* as well.[1,2] There is a slight shift in the diffraction peaks of the obtained patterns towards smaller *d*-values. The sample obtained in experiment III using Pb$_2$OSO$_4$, CuO and P$_{red}$ resulted in an excellent phase with only traces of Cu impurity and some PbS crystals on cold end of the tube. Visibly inspection under microscope revealed this phase contained of only dark grey lumps intermixed with some transparent (or white) material, where as other two phase (I and II) also contained rust colored and or yellowish substance present in significant amount. The X-ray patterns of other samples are shown in figure 2. Notably, the Pb-oxyapatite phase is only formed if Pb-sulphates (PbSO$_4$ or Pb$_2$OSO$_4$) are used as a precursor in various combinations with other starting materials. Thus, the use of Pb-sulphate appears to be very crucial for the synthesis of the LK-99 type phase. For all the experiments where Pb source was not the sulphate a different from LK-99 resulted, most commonly the monoclinic Pb$_3$(PO$_4$)$_2$[11] or the Pb$_8$(PO$_4$)$_2$ phase[3] (Figure 3 and 4). This shows that it is quite difficult to synthesize a phase pure material by using a different combination of stoichiometric starting materials

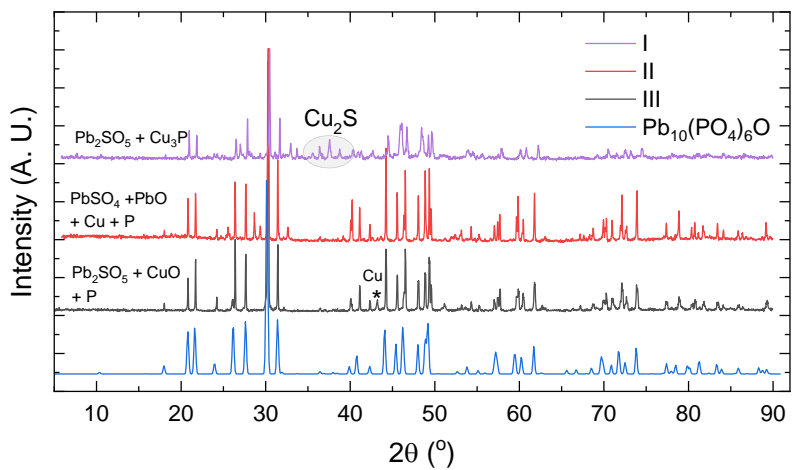

Figure 2. PXRD patterns of 'LK-99' phase obtained from synthetic route I-III (black, red and purple lines) in comparison to the calculated powder pattern of $Pb_{10}(PO_4)_6O$ (bottom most, blue line) compound.

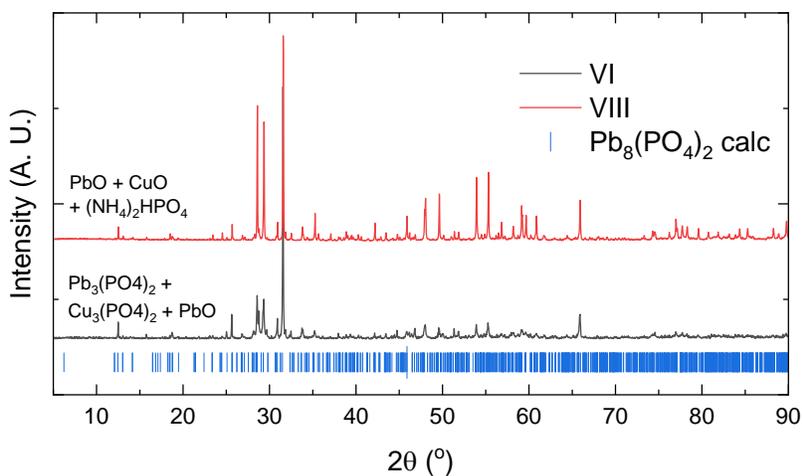

Figure 3. PXRD patterns of various products obtained from synthetic route IV and V (bottom black and top red lines). Vertical blue bars are the allowed Bragg reflections for monoclinic $Pb_3(PO_4)_6$ phase.

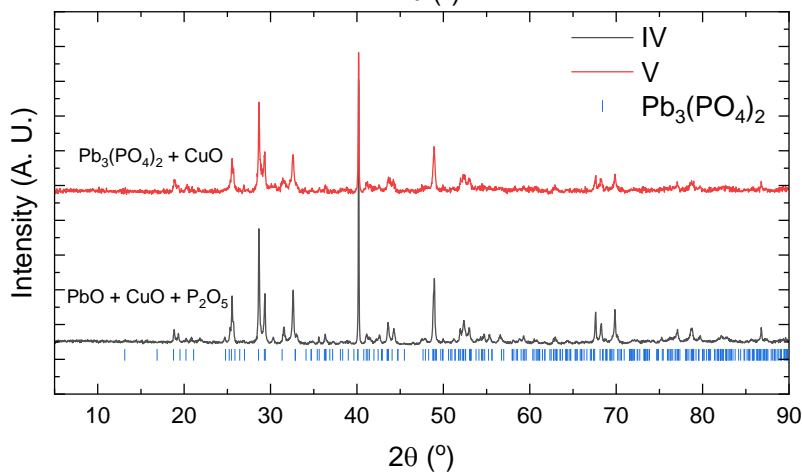

Figure 4. PXRD patterns of various products obtained from synthetic route VI and VII (bottom black and top red lines). Vertical blue bars are the allowed Bragg reflections for monoclinic $Pb_8(PO_4)_2$ phase.

We shall now only focus on the products that yielded the LK-99 type phase. Although the PXRD appears to be very clean for at least one sample, a visual examination of the samples under optical microscope clearly shows the multiphase nature of the samples. Along with dark greyish-looking lumps, there are regions, which are pale-yellowish transparent or opaque rust-colored (figure 5), depending on the reactants used. An inhomogeneous and multiphase product is naturally expected for a sample prepared by synthesis route (I), the original route to obtain the LK-99 phase reported by Lee *et al*[1,2]. The following balanced chemical equation for the reaction pathway that Lee *et al* followed suggests the products and their respective amounts.

$$2\ Pb_2(SO_4)O + 2\ Cu_3P \rightarrow 1/3\ Pb_9Cu(PO_4)_6O + 2\ Cu_2S + 5/3\ CuO + Pb$$

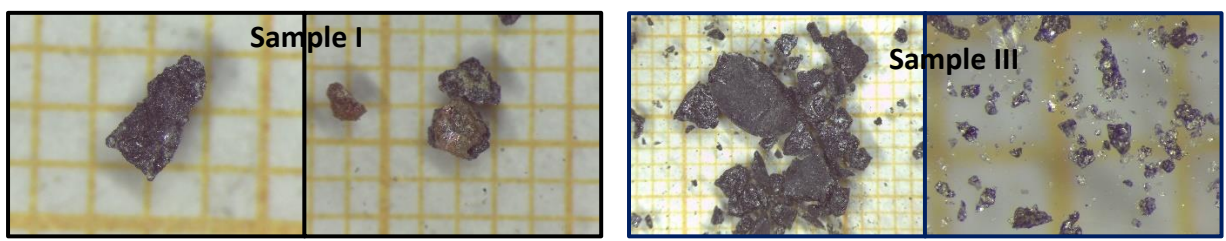

Figure 5. Optical images of the different 'LK-99' phases obtained on mm grid. For sample I, a rust-colored and yellowish mass can be seen in the right picture. Sample III contains transparent pale-yellowish crystal next to a black crystalline mass.

According to the above equation, $Cu_2S$ must appear as a significant impurity along with the LK-99 phase, which is clearly observed in PXRD pattern of Lee *et al*'s. Other authors, Awana *et al,*[4] and Jia *et al*[6] also encountered this impurity. For the PXRD data of the best sample, we chose to characterize only the dark grey appearing part of the sample, disregarding the rest. This results in a clean looking PXRD pattern. Most recent reports by Awana *et al*[5] and Shi et al[8] showed a very clean looking PXRD pattern suggesting a (nearly) 'phase-pure' sample, which we believe is impossible to achieve if characterization is performed on whole sample. However, they did not describe the process of sample preparation for PXRD measurements. The sample selection process has greater implication on the composition of the target phase as it can influence the determination of correct (idealized) composition 'Pb$_9$Cu(PO$_4$)$_6$O'. The presence of significant impurities in the whole sample might indicate a deviation of the actual composition from the true ideal composition leading to different (observable) properties. We will try to clarify this in our later compositional analysis and magnetic properties sections.

**Magnetic levitation**

The three samples that showed the purported superconducting LK-99 phase were checked for magnetic levitation as a test for diamagnetism associated with RTSC. Small chunks were placed over a strong Nd-magnet but no visible levitation or repulsion was observed. However, in some sample batches (I, II and VII) we observed few tiny flakes responded to a permanent magnet. In the presence of a magnet, the flakes appear to align vertically and tilt to the direction opposite to the movement of the magnet; however, the particles themselves were not attracted towards the magnet. This observation gave an impression of 'half-levitation' (Figure 6) by the particles. Similar behavior was also recently observed by Chang et al[9] interpreting this as magnetic levitation due to superconductivity. It is worth pointing out that no elemental analysis was performed on the levitating particle by these authors to confirm if the particle has the correct LK-99 composition. However, we did an elemental analysis of many of these particles and found that some of them contain traces of iron on the surface while others do not. This is the likely reason the particles respond to the magnet at least in the former case. This, Fe impurity is present in extremely trace amount and its origin is not identified yet. The most likely source of Fe contamination is the use of metallic spatula for weighing the sample (the sample was scraped from mortar pestle using a plastic spatula though). Hence, any such observation must be interpreted cautiously and not be confused with the Meissner effect or levitation due to superconductivity until other strong experimental evidences suggest so. Until we perform a more detailed analysis, it is safe to assume that the 'half-levitation' arises due to some trace ferromagnetic impurity. Li *et al*[6] have also observed similar behavior in their sample and have thoroughly investigated it through magnetic studies. They indeed see a soft ferromagnetic response in their magnetic data confirming that the half-levitation observed is not due to superconductivity but related to the sample. However, even in their study a compositional analysis of such levitating particle is lacking. Similar weak soft ferromagnetic behavior is observed in our samples too, which is discussed in later section.

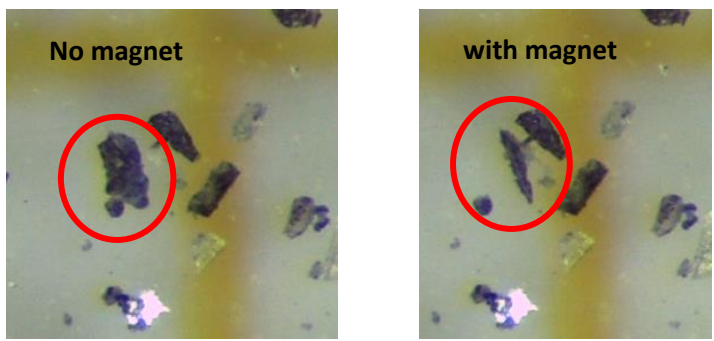

Figure 6. Tiny flakes of sample aligning vertically after placing a permanent magnet underneath.

**Compositional Analysis:**

Lee *et al*[1,2] emphasized the importance of Cu-doping to achieve superconductivity in the LK-99 phase. Hence, it is essential to investigate the actual amount of copper substituted in the sample. We performed semi-quantitative elemental analysis using SEM-EDX on all the 'LK-99' phases we obtained (Figure 7). All the regions in the tested samples show the presence of Pb, P and O in the desired ratio. However, no copper (or only a trace amount of it in some regions) could be detected in all the three samples. This finding indicates that the samples are microscopically inhomogeneous too and phase separation occurs within the black (or dark grey) looking grains. Apart from the desired phase, we detected crystals of PbS embedded within the grains of the Pb-oxyapatite phase and on the surface of the crystals, resulting in a slightly higher Pb:P ratio of ~ 2 in some regions (Figure 7a). $Cu_2S$ was clearly identified in many regions as tiny crystals sitting on top of larger Pb-oxyapatite crystallites (Figure 7b). To confirm the presence, or rather the absence, of Cu in the Pb-oxyapatite phase we did elemental mapping of some regions of sample I (result of one such region is shown in Figure 7). It is clear from the elemental mapping that copper is present but largely on the surface of the sample as $Cu_2S$ impurities and only in trace amount in the in the desired phase. Surprisingly all the reports [4, 5, 7-9] except one on replication of Lee et al,[1,2] work lack any compositional analysis. Only Jia et al[6] have shown the EDX analysis of their LK-99 sample, which indeed shows the presence of copper in one region. However, it is not clear how much copper is actually substituted. In their EDX spectrum, S is also detected in large quantities, which raises the possibility that the detected Cu actually comes from the $Cu_2S$ phase on the surface and is not be incorporated in the Pb-oxyapatite phase. Chemical intuition also supports this conjecture, as the ionic radii[12] and coordination preferences of $Pb^{2+}$ and $Cu^{2+}$ are significantly different, as well as their chemical behavior. Therefore, it is challenging to substitute copper appreciably at Pb-sites and this might partly be the reason why so much excess copper is required to obtain the desired phase in case where the SC LK-99 phase does form. It highlights the different chemical environments preferred by the two ions. It is emphasized by Lee et al that Cu-doping is highly crucial for superconductivity in the LK-99 phase; however, they have not presented any elemental analysis that could reveal the correct stoichiometry of the superconducting compound. Shrinkage of the lattice parameter was taken as the only evidence of Cu-doping, which may occur for reasons other than this. Without any copper doping the sample ($Pb_{10}(PO_4)_6O$) should be insulating and diamagnetic, which other authors also observed.[5,6] Knowledge of the exact amount of copper in the superconducting phase is a crucial for reliably reproducing the RTSC 'LK-99' phase and should be presented to consolidate the claim of SC in Cu-doped $Pb_{10-x}Cu_x(PO_4)_6O$ compound.

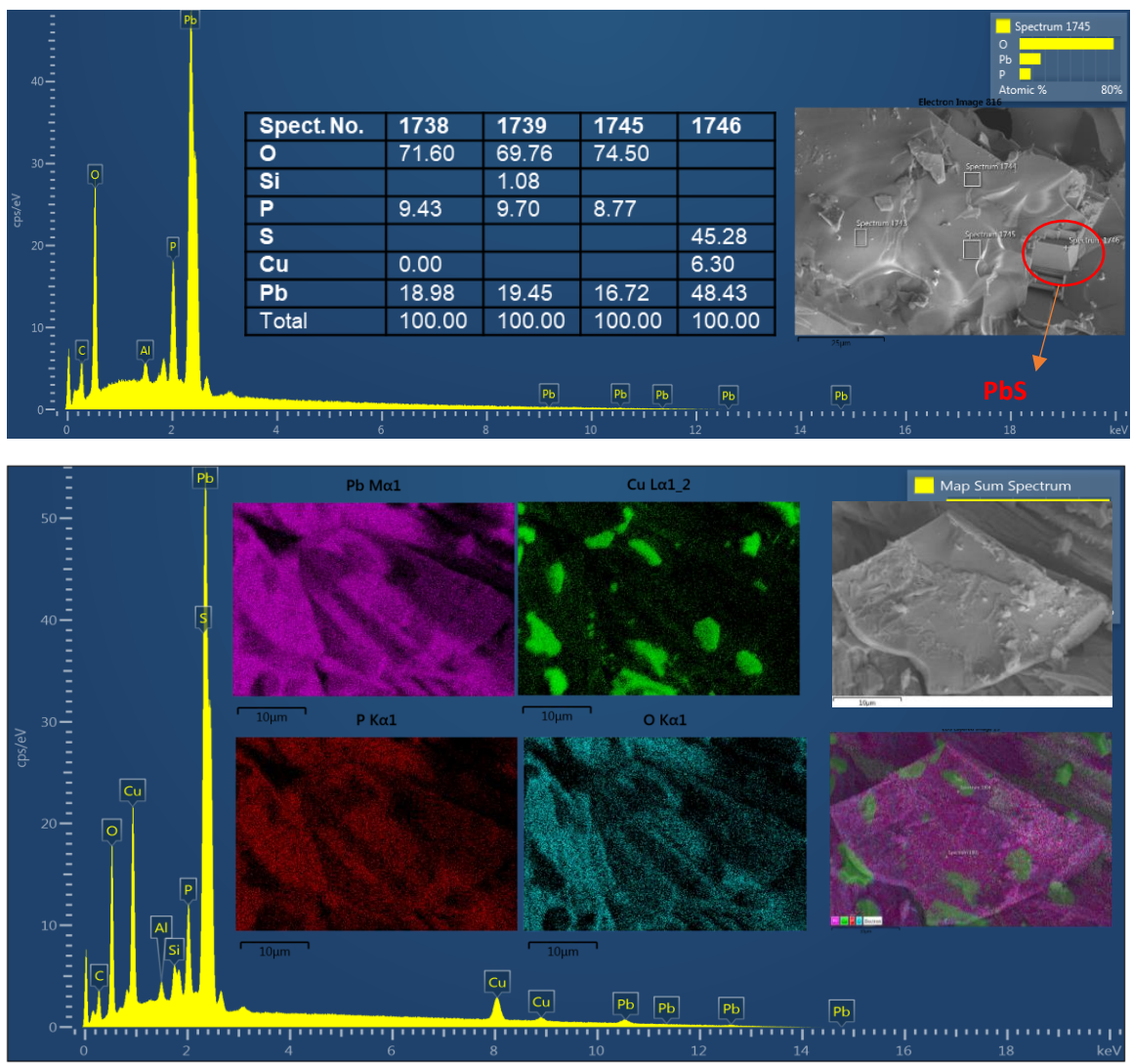

Figure 7. (a) Representative EDX spectrum and the secondary electron image of a region of sample obtained from synthetic route III (b) EDX spectrum, secondary electron image and elemental mapping of a representative crystal of sample obtained from route I. Silicon in both spectra arise from the surface contamination from the quartz tube. Peaks for Al and C arise from the stub and carbon tape used to mount the sample.

**Magnetic properties.**

Magnetic data on two independent samples were collected using temperature-dependent and isothermal magnetization measurements. Both the samples were found to be diamagnetic with no perceptible temperature dependence (Figure 8) or divergence in ZFC/Fc data. Surprisingly the low-temperature hysteresis curves (Figure 8 (b and d)) indicates a weak and soft ferromagnetic-like behavior, which could partly explain the half-levitation phenomena. However this is quite intriguing as no 'magnetic elements' (containing $d$ electrons) are present in the sample and due care was taken not to use any metal spatula for weighing or scraping the sample powder while preparation of sample (I). Even at 325 K, this FM-like behavior persists but is primarily superimposed by the strong diamagnetic response of the sample. Jia *et al*[6] observed similar features in their sample and Awana *et al*[4] also obtained a similar hysteresis curve at 280 K in their first sample. We want to point out that the magnitude of the saturation magnetization (~5.5 × $10^{-2}$ emu g$^{-1}$ at 100 and 300 K) obtained by Jia *et al* nearly matches with that of our sample I (~5.7× $10^{-2}$ emu g$^{-1}$) which may exclude the possibility of a magnetic impurity. We wonder if this is intrinsically related to the sample or is an impurity effect that occurs independently in multiple samples studied by from different groups. Clearly, further detailed analysis is needed to understand such a strange behavior.

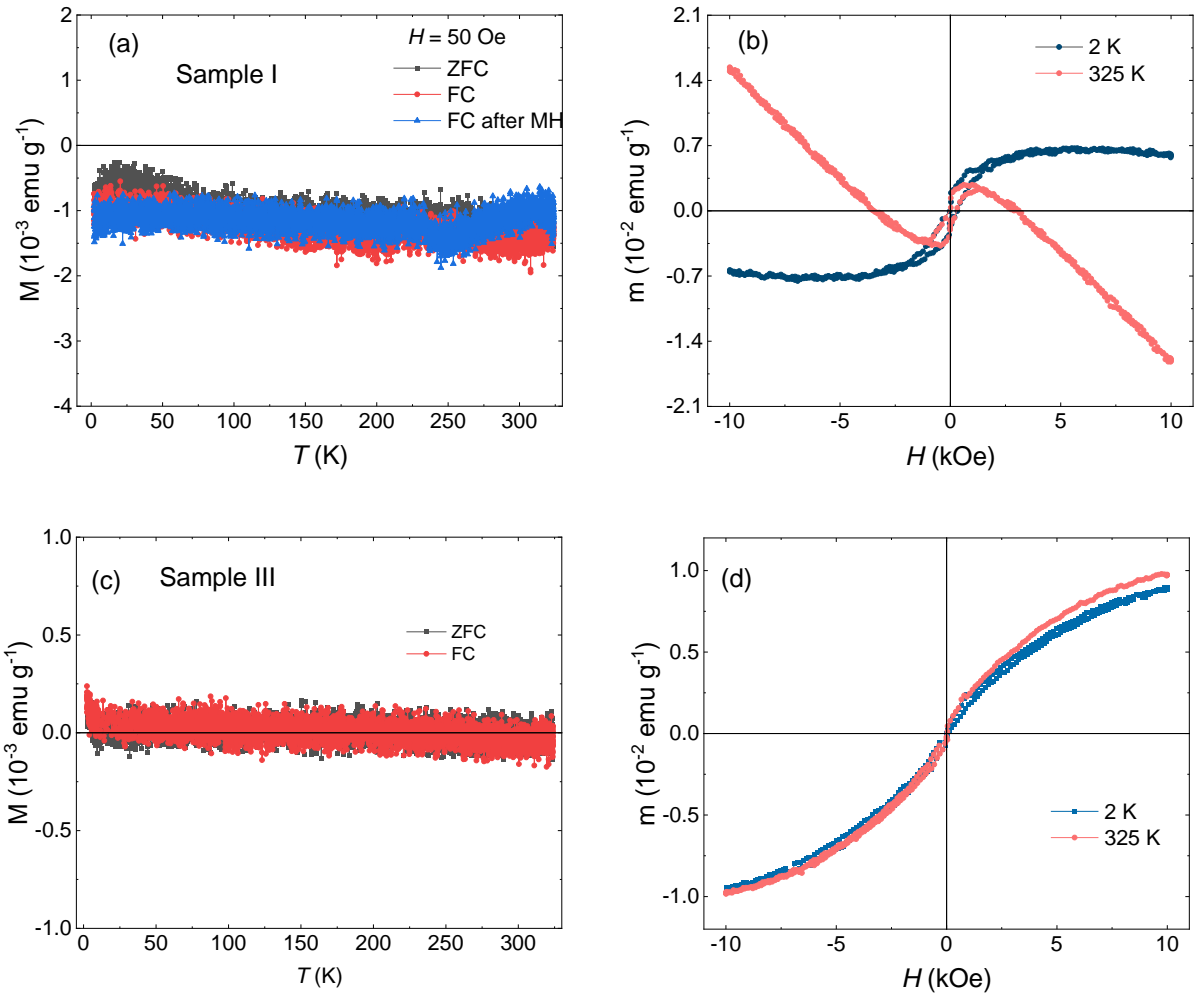

Figure 8. (a) Temperature dependent and (b) isothermal magnetization for LK-99 sample I and (c) Temperature dependent and (d) isothermal magnetization for LK-99 sample III.

**Conclusions and outlook**

Through our elaborate experimental work, we conclude that it is rather challenging to achieve the Pb-oxyapatite phase by solid-state method and only using $PbSO_4$ or $Pb_2OSO_4$ results in the formation of the desired phase. The samples appear diamagnetic but superconductivity or magnetic levitation cannot be confirmed in any of the LK-99 phases we obtained. A notable conclusion from our results is that incorporation Cu at Pb sites proves to be extremely difficult, if not impossible, to accomplish as most of our samples did not or contained only trace amounts of Cu in the desired phase. This is supported by the fact that the ionic radii of $Pb^{2+}$ and $Cu^{2+}$ differ as significantly as d their chemistry. We urge other authors to conduct more rigorous compositional and structural analysis on their samples to ascertain, if at all, how much copper is actually substituted at the Pb sites. This would also help in identifying the optimum superconducting composition and possibly isolating a phase pure material by alternative routes. Until then a large uncertainty looms over the true structure and composition of the superconducting phase. As Lee *et al* stressed that Cu-doping is essential for observing RTSC in this phase, the absence of SC in the samples reproduced so far might actually be related to insufficient copper doping! If Cu-doping is indeed, the key to achieving RTSC in this material then sample preparation using high-pressure technique or hydrothermal growth might prove beneficial in stabilizing the Cu-doped phase. Efforts to grow the LK-99 phase using hydrothermal method are underway.

**Acknowledgement**

GST acknowledges the financial support from 'Würzburg-Dresden Cluster of Excellence,' *ct.qmat*.